\documentclass[5p,twocolumn]{elsarticle}  

\usepackage{graphicx}  
\usepackage{graphics}
\usepackage{dcolumn}   
\usepackage{bm}        
\usepackage{amssymb}   
\usepackage{amsmath}   
\usepackage{caption}
\usepackage{epsfig}
\usepackage{color}

\begin{document}
\begin{frontmatter}

\title{
Virtual Compton Scattering measurements in the nucleon resonance
region}

\author[temple]{A.~Blomberg}
\address[temple]{Temple University, Philadelphia, PA 19122, USA}

\author[temple]{H.~Atac}

\author[temple]{N.~Sparveris\corref{cor1}}
\ead{sparveri@temple.edu} \cortext[cor1]{Corresponding author.}

\author[temple]{M.~Paolone}

\author[mainz]{P.~Achenbach}
\address[mainz]{Institut f\"ur Kernphysik, Johannes Gutenberg-Universit\"at Mainz, D-55099 Mainz, Germany}

\author[cfrnd]{M.~Benali}
\address[cfrnd]{Clermont Universite, Universite Blaise Pascal,
CNRS/IN2P3, LPC, BP 10448, F-63000 Clermont-Ferrand, France}

\author[ijs]{J.~Beri\v{c}i\v{c}}
\address[ijs]{Jo\v{z}ef Stefan Institute, SI--1000 Ljubljana, Slovenia}

\author[mainz]{R.~B\"ohm}
\author[mainz]{L.~Correa}
\author[mainz]{M.O.~Distler}
\author[mainz]{A.~Esser}
\author[temple]{D.~Flay}
\author[cfrnd]{H.~Fonvieille}
\author[mainz]{I.~Fri\v{s}\v{c}i\'{c}}
\author[mainz]{Y.~Kohl}
\author[mainz]{H.~Merkel}
\author[mainz]{U.~M\"uller}
\author[temple]{Z.~E.~Meziani}
\author[ijs]{M.~Mihovilovic}
\author[mainz]{J.~Pochodzalla}
\author[temple]{A.~Polychronopoulou}

\author[pavia]{B.~Pasquini}
\address[pavia]{Dipartimento di Fisica, Universit\`{a} degli Studi di Pavia, I-27100 Pavia, Italy}

\author[mainz]{M.~Schoth}
\author[mainz]{F.~Schulz}
\author[mainz]{S.~Schlimme}
\author[mainz]{C.~Sfienti}
\author[ijs,ul]{S.~Sirca}
\address[ul]{Faculty of Mathematics and Physics, University of Ljubljana, SI--1000 Ljubljana, Slovenia}

\author[mainz]{A.~Weber}


\begin{abstract}
We report on new measurements of the electric Generalized
Polarizability (GP) of the proton $\alpha_E$ in a kinematic region
where a puzzling dependence on momentum transfer has been observed,
and we have found that $\alpha_E = (5.3 \pm 0.6_{stat} \pm
1.3_{sys})~10^{-4} fm^3$ at $Q^2=0.20~(GeV/c)^2$. The new
measurements, when considered along with the rest of the world data,
suggest that $\alpha_E$ can be described by either a local plateau
or by an enhancement in the region $Q^2=0.20~(GeV/c)^2$ to
$0.33~(GeV/c)^2$. The experiment also provides the first measurement
of the Coulomb quadrupole amplitude in the $N \rightarrow \Delta$
transition through the exploration of the $p(e,e'p)\gamma$ reaction.
The new measurement gives $CMR = (-4.4 \pm 0.8_{stat} \pm
0.6_{sys})~\%$ at $Q^2=0.20~(GeV/c)^2$ and is consistent with the
results from the pion electroproduction world data. It has been
obtained using a completely different extraction method, and
therefore represents a strong validation test of the world data
model uncertainties.
\end{abstract}


\end{frontmatter}



\section{Introduction}

The polarizabilities of a composite system such as the
nucleon~\cite{reviewpolar} are fundamental structure constants, just
as its size and shape, and can be accessed experimentally by Compton
scattering processes. In the case of real Compton scattering (RCS),
the incoming real photon deforms the nucleon, and by measuring the
energy and angular distributions of the outgoing photon one can
determine the global strength of the induced current and
magnetization densities, which are characterized by the nucleon
polarizabilities. Although the electric, magnetic and some of the
spin polarizabilities are known with reasonable accuracy from
Compton scattering experiments, little is known about the
distribution of polarizability density inside the nucleon. We can
gain access to this information through the virtual Compton
scattering (VCS) process where the incident real photon is replaced
by a virtual photon~\cite{Guichon:1998xv}. The virtuality of the
photon allows us to map out the spatial distribution of the
polarization densities. In this case it is the momentum of the
outgoing real photon $q'$ that defines the size of the perturbation
while the momentum of the virtual photon $q$ sets the scale of the
observation. In analogy to the form factors for elastic scattering,
which describe the charge and magnetization distributions, VCS gives
access to the deformation of these distributions under the influence
of an electromagnetic field perturbation as a function of the
distance scale. The structure dependent part of the process is
parametrized by the Generalized Polarizabilities (GPs) which can be
seen as Fourier transforms of local polarization densities
(electric, magnetic, and spin)~\cite{Gorchtein:2009qq}. The GPs are
therefore a probe of the nucleon dynamics, allowing us, e.g., to
study the role of the pion cloud and quark core contributions to the
nucleon dynamics at various length scales.

The GPs depend on the quantum numbers of the two electromagnetic
transitions involved in the Compton process and typically a
multipole notation is adopted. Initially ten independent
lowest-order GPs were defined~\cite{gp2}; it was
shown~\cite{gpred1,gpred2} that nucleon crossing and charge
conjugation symmetry  reduce this number to six, two scalar ($S=0$)
and four spin, or vector GPs ($S=1$). The two scalar GPs, the
electric and the magnetic, generalize the well known static electric
$\alpha_E$ and magnetic $\beta_M$ polarizabilities obtained in real
Compton scattering \cite{gp1,rcs}. Contrary to atomic
polarizabilities, which are of the size of the atomic volume, the
proton electric polarizability $\alpha_E$ \cite{rcs} is much smaller
than the volume scale of a nucleon (only a few \% of its volume).
The small size of the polarizabilities reveals the extreme stiffness
of the proton as a direct consequence of the strong binding of its
inner constituents, the quarks and gluons, while representing a
natural indication of the intrinsic relativistic character of the
nucleon. In most recent theoretical models the electric GP
$\alpha_E$ is predicted to decrease monotonically with $Q^2$. The
observed smallness of $\beta_M$ relative to $\alpha_E$ can be
explained by the existence of the competing paramagnetic and
diamagnetic contributions, which nearly cancel. Furthermore, the
$\beta_M$ polarizability is predicted to go through a maximum before
decreasing. This last feature is usually explained by the dominance
of diamagnetism due to the pion cloud at long distance, or small
$Q^2$, and the dominance of paramagnetism due to a quark core at
short distance (large $Q^2$).


VCS is accessed experimentally by exclusive photon
electroproduction. The main kinematic variables are the CM 3-momenta
$\vec{q}_{cm}$ and $\vec{q}^{'}_{cm}$ of the initial and final
photons respectively, the CM angles of the outgoing real photon
w.r.t. $\vec{q}_{cm}$: the polar angle $\theta_{\gamma^*\gamma}$ and
the azimuthal angle $\phi$, and the 4-momentum transfer squared
$Q^2$. The photon electroproduction amplitude can be decomposed into
the Bethe-Heitler (BH), the Born, and the non-Born contributions.
The BH and VCS Born parts are well known and entirely calculable
with the nucleon electromagnetic (EM) form factors as inputs, while
the non-Born part involves the structure dependent component,
parametrized by the GPs. In order to extract the GPs from
measurements of photon electroproduction cross sections one can
utilize two methods. The first method involves the Low Energy
Theorem (LET)~\cite{gp2} and is valid below pion threshold only. The
second method involves the Dispersion Relations (DR)
approach~\cite{gp12,gp13,Pasquini:2018wbl} and its domain of
validity includes the $\Delta(1232)$ resonance up to the $N\pi\pi$
threshold. When the VCS reaction is measured in the $\Delta(1232)$
resonance region the VCS non-Born part offers access to additional
information, such as the $N \rightarrow \Delta$ transition form
factors~\cite{vcspaper}. An extensive experimental and theoretical
effort has focused on this subject in the past two decades.
Particular attention has been addressed to the two quadrupole
transition amplitudes, the Electric and the Coulomb, which offer a
path for the exploration of the non-spherical components in the
nucleon wavefunction. A review of this topic is presented
in~\cite{soh}.

\begin{figure}[!]
\includegraphics[width = \columnwidth]{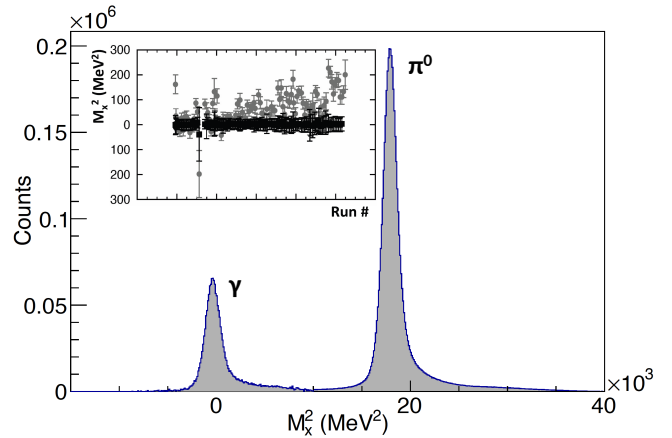}
\caption{The missing mass spectrum. The two peaks corresponding to the photon and to the $\pi^0$ are very well separated. The photon peak has been
multiplied by a factor of 10 so that it can be clearly seen in the
figure. The inserted panel shows the center of the photon missing
mass peak before (gray circle) and after (black box) the momentum
calibration as a function of the different run numbers.} \label{fig:tofmm}
\end{figure}

\section{The experimental measurements}

The measurements were performed at $Q^2=0.20~(GeV/c)^2$ and in the
$\Delta(1232)$ resonance region. The experiment aims to study the
electric Generalized Polarizability of the proton $\alpha_E$ in a
region where two MAMI experiments~\cite{gp3,gp4} have identified an
unexpected enhancement of $\alpha_E$, at $Q^2=0.33~(GeV/c)^2$, that
can not be accommodated along with the other experimental
measurements at different $Q^2$ with a single dipole fall-off in
$Q^2$. In this work an additional opportunity is also presented to
access for the first time the $N \rightarrow \Delta$ quadrupole
amplitudes through the measurement of the photon channel and to
offer a valuable cross check to the world data. The two quadrupole
transition amplitudes have so far been explored only through pion
electroproduction measurements. In this work we have explored the
Coulomb quadrupole for the first time through the VCS reaction,
providing a measurement through a different reaction mechanism and
within a completely different theoretical framework.

The experiment kinematics focused on $\theta_{\gamma^*\gamma} >
120^\circ$ in order to avoid the kinematic region where the BH
process dominates and to allow for the sensitivity to the measured
signal to be maximal. The VCS cross section was measured, as well as
the in-plane azimuthal asymmetry of the VCS cross section with
respect to the momentum transfer direction
\qquad \\
$$A_{(\phi_{\gamma^*\gamma}=0,\pi)} =
\frac{\sigma_{\phi_{\gamma^*\gamma}=0} - \sigma_{\phi_{\gamma^*
\gamma}=180}}  {\sigma_{\phi_{\gamma^*\gamma}=0} +
\sigma_{\phi_{\gamma^*\gamma}=180}}.$$
\qquad \\
The asymmetry allows for part of the cross section's systematic
uncertainties to be suppressed thus improving the precision of the
measurements and amplifying the sensitivity to the measured
amplitudes.

The experiment utilized the A1 spectrometer setup at
MAMI~\cite{mamisetup}. An 1.1~GeV unpolarized electron beam with a
beam current of $~20\mu A$ was employed on a 5~cm liquid hydrogen
target, while the beam was rastered across the target to avoid
boiling. The recoil proton and the scattered electron were detected
in coincidence with spectrometers A and B~\cite{mamisetup}, while
the undetected photon of the VCS reaction was identified through the
missing mass spectrum. Each spectrometer  contains two vertical
drift chambers, two scintillator planes, and a Cherenkov detector,
while both spectrometers offer a momentum resolution of $10^{-4}$.
The experimental setup offered a better than 1~ns (FWFM) resolution
in the coincidence timing spectrum, while the subtraction of the random coincidences introduced a minor uncertainty as a result of the small contribution of the
random coincidences under the coincidence timing peak and of the
excellent timing resolution. The photon electroproduction events
were further identified through the missing-mass reconstruction as
shown in Fig.~\ref{fig:tofmm}~(right panel), where one can clearly observe the
photon peak as well as the $\pi^{\circ}$-electroproduction peak
which also falls within the spectrometer acceptance, with the two
peaks being very well separated. An extensive description of the
experimental arrangement and parameters, as well as of the data
analysis can be found in~\cite{blombergthesis,mamiproposal}.

A calibration was performed to the central momenta of the two
spectrometers by simultaneously optimizing the experimental missing
mass peak position and width. An additional constraint in determining the actual momentum settings of both spectrometers is offered by the fact that in the asymmetry
measurements the electron spectrometer settings as well as the proton spectrometer momentum remain fixed~\cite{blombergthesis}. The calibration
resulted in a 0.4\% correction to the proton spectrometer momentum setting and in a 0.1\% to the electron
spectrometer one. In Fig.~\ref{fig:tofmm}~(right panel insert) one can see the
center of the photon missing mass peak before and after the momentum
calibration as a function of the different run numbers.


\begin{figure}[!]
\includegraphics[width = \columnwidth]{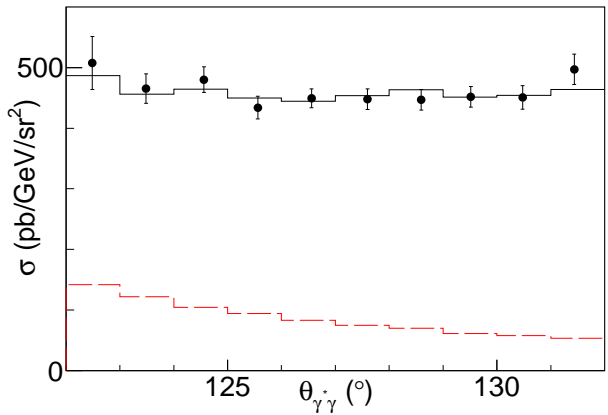}
\caption{Experimental cross sections as a function of the
spectrometer acceptance at $\phi_{\gamma^*\gamma}=0^\circ$. The BH+Born contribution (dashed line) is
compared to the DR calculation for the total cross section (solid
line).} \label{fig:sigmasimul}
\end{figure}

\begin{figure*}[t]
\centering \includegraphics[width = 16.0cm]{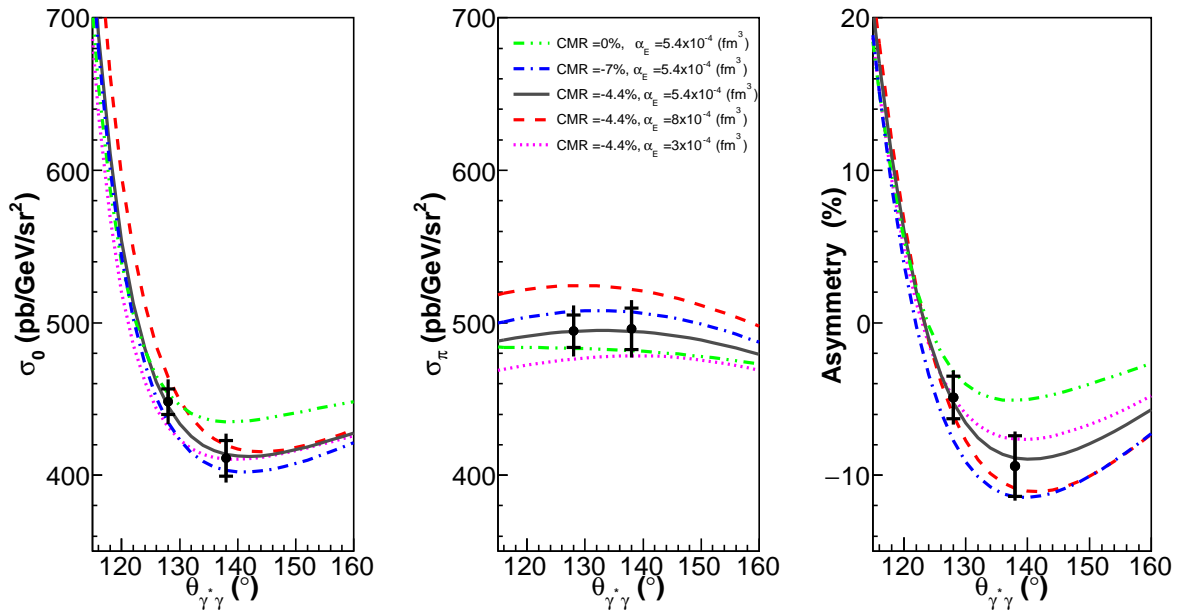}
\caption{Cross sections and asymmetries measured at
$Q^2=0.20~(GeV/c)^2$. The solid line corresponds to the DR
calculation with the extracted values for $\alpha_E$ and for $CMR$. The
dashed and the dotted curves show the effect from a variation to the
central value of $\alpha_E$, from $3~10^{-4} fm^3$ to $8~10^{-4}
fm^3$, while the dash-dot-dot and the dash-dot curves show the effect
from the variation to the $CMR$ value, from 0$\%$ to
-7$\%$.} \label{fig:sigma}
\end{figure*}

In order to determine the experimental cross section the five-fold
solid angle is calculated through a Monte Carlo simulation that
offers an accurate description of the experimental setup, including
the intrinsic resolution of the detectors and energy losses. The
radiative corrections have been included in the simulation and they
are accounted for as described in~\cite{radiative}. In order to
extract the experimental cross section the generated events in the
simulation are weighted with a cross section using the DR
calculation of \cite{gp12,gp13}. The calculation includes the
BH+Born and the non-Born contributions, where one can utilize
different parameterizations for the non-Born part. For the non-Born
part a realistic initial parametrization is applied based on the
current knowledge of the GPs as well as that of the $N \rightarrow
\Delta$ transition amplitudes, the experimental cross sections are
determined, and the scalar GPs are then set as free parameters and
are extracted utilizing the DR
framework~\cite{blombergthesis,jlabgp}. Then the process is repeated
by using the extracted parameters as a new input in the simulation
cross section, and the amplitudes of interest are then extracted
again. This procedure converges quickly and the extracted values for
the GPs are at that point finalized. If the procedure is repeated by
utilizing a different initial parametrization for the non-Born part
the converging results are independent of the initial input
parameters. The BH+Born contribution accounts for $\approx 20\%$ of
the total cross section (see Fig.~\ref{fig:sigmasimul}) for all
experimental settings. The primary sources of systematic
uncertainties involve the uncertainties in the momenta and the
angles of the two spectrometers, the luminosity, the knowledge of
the acceptance, and the radiative corrections. For the spectrometer
momenta the effect of an uncertainty of $\pm 2~10^{-4}$ was studied
by varying the momentum settings accordingly, repeating the analysis
process, and the deviation of the extracted results was quantified
as the corresponding uncertainty. A similar procedure was also
followed to study the effect of the uncertainty in the spectrometer
angles, where the variation involved was $\pm 0.1~mr$ for each one
of the two spectrometers. The effect of these uncertainties varies
among the settings but is in principle of the order of $\pm 1\%$.
The uncertainty to the solid angle, the luminosity, and the
radiative corrections added quadratically is $\approx \pm 2.5 \%$.
The statistical uncertainty on the cross section is typically
smaller, ranging between $1.5 \%$ and $2 \%$.

For the electric Generalized Polarizability the extracted value is
$\alpha_E = (5.3 \pm 0.6 \pm 1.3)~10^{-4} fm^3$ at
$Q^2=0.20~(GeV/c)^2$, where the first uncertainty is statistical and
the second is the systematic one. The magnetic GP was also treated
as a free parameter in the analysis but was constrained within a
very large uncertainty, something to be expected since the
experiment kinematics were optimized for the measurement of
$\alpha_E$. The effect of using different parameterizations for the
proton form factors in the analysis was explored since these
quantities enter the calculation of the BH+Born cross section. A
systematic study was performed by applying different
parameterizations in the analysis, and the variation of the
$\alpha_E$ results was determined to be $\pm 0.3~10^{-4} fm^3$. This
value was treated as a systematic uncertainty and was integrated
into the final systematic uncertainty.

The experimental measurements offer sensitivity to the $N
\rightarrow \Delta$ transition Coulomb quadrupole amplitude, which
is typically quantified through the $CMR$ ratio to the dominant
magnetic dipole amplitude. The Coulomb quadrupole has been
previously measured at the same $Q^2$ through the $p(\vec{e},e'p)\pi^0$ reaction,
utilizing the same experimental setup as in this work, giving the
result $CMR = (-5.09 \pm 0.28_{stat+sys} \pm
0.30_{mod})~\%$~\cite{spaplb}. If this amplitude is set as a free
parameter in the analysis of the VCS measurements we derive the
value of $CMR = (-4.4 \pm 0.8_{stat} \pm 0.6_{sys})~\%$, which is in
very good agreement with the result derived from the pion channel
measurement~\cite{spaplb}. In this case the central value for the
electric GP increases slightly to $\alpha_E = 5.4~10^{-4} fm^3$. In
Fig.~\ref{fig:sigma} the measured cross sections and asymmetries for
a fixed $Q^2=0.20~(GeV/c)^2$ are presented. The solid line
corresponds to the DR calculation with the extracted values for
$\alpha_E$ and $CMR$ as an input. The dashed and the dotted lines
exhibit the effect from the variation of $\alpha_E$, while the
dash-dot and the dash-dot-dot lines show the effect of a variation
on the $CMR$.



\begin{figure*}[t]
\centering \includegraphics[width = 16.0cm]{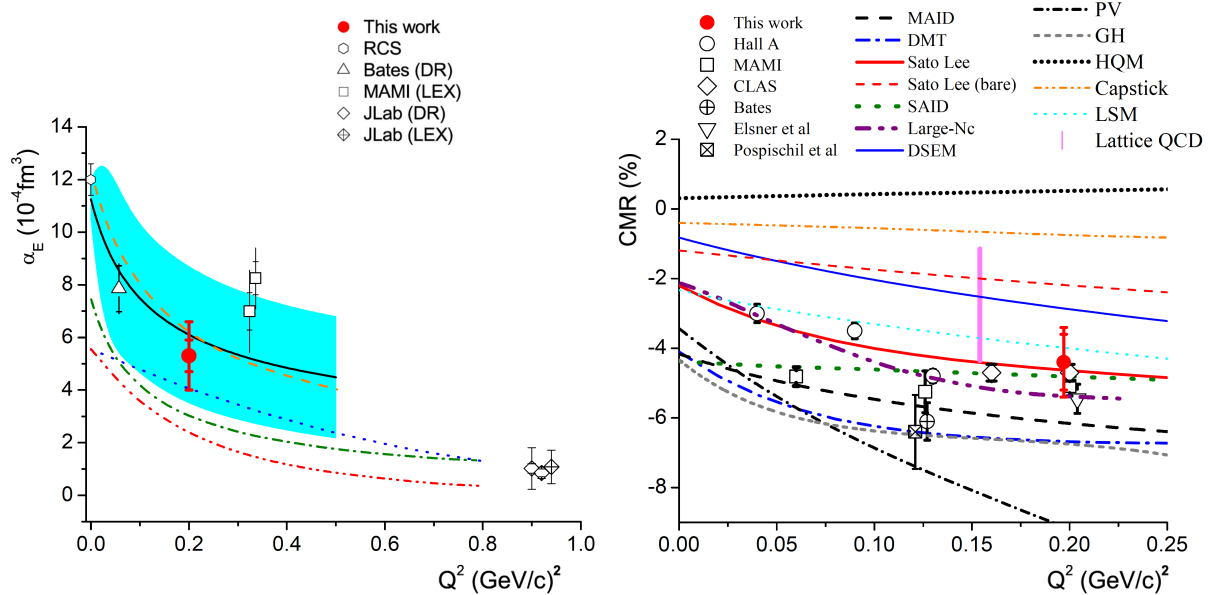}
\caption{Left panel: The world data on the electric GP, with
statistical (inner) and total error bar. The result from this work
(solid circle) as well as the measurements from Bates~\cite{gp6},
MAMI~\cite{gp3,gp4,helprog}, JLab~\cite{gp5,jlabgp} and the
RCS~\cite{rcs} are shown. The theoretical calculations
BChPT~\cite{bchpt} (solid line with uncertainty band),
HBChPT~\cite{gp9} (dash), NRQCM~\cite{gp15} (dot), Effective
Lagrangian Model~\cite{gpkorchin} (dash-dot-dot) and Linear Sigma
Model~\cite{gp16} (dash-dot) are also shown. Right panel: The CMR
measurement from this work (solid circle) and from
\cite{spaplb,pos01,spaprl,stave,elsner,aznauryan,ndeltaplb,sparepja}
(open symbols) are presented. All data points are shown with their
total experimental
 uncertainties (statistical and systematic) added in quadrature. The theoretical
predictions of MAID \cite{kama,mai00}, DMT \cite{dmt}, SAID
\cite{said}, Sato-Lee \cite{sato_lee}, Capstick \cite{capstick}, HQM
\cite{hqm}, the Lattice-QCD calculation \cite{dina}, the large-Nc
calculation \cite{largenc}, the DSEM~\cite{dsem}, the linear
$\sigma$-model (LSM)~\cite{lsm}, the ChEFT  of
Pascalutsa-Vanderhaegen (PV)~\cite{pasc} and the Gail-Hemmert (GH)
\cite{hemmert} are also shown.} \label{fig:results}
\end{figure*}

\section{Discussion and conclusions}

In this experiment we have performed a measurement of $\alpha_E$ in
a kinematic region where a puzzling dependence on the momentum
transfer has been observed. The result for the electric GP is shown
in Fig.~\ref{fig:results} (left panel), along with the world data
from Bates~\cite{gp6}, MAMI~\cite{gp3,gp4}, JLab~\cite{gp5,jlabgp}
and the RCS~\cite{rcs}. The new result suggests that $\alpha_E$
follows a fall-off with $Q^2$ after the Bates data point~\cite{gp6}
at $Q^2=0.06~(GeV/c)^2$. Then $\alpha_E$ can be described by either
an enhancement, or by a local plateau, in the range of
$Q^2=0.20~(GeV/c)^2$ to $0.33~(GeV/c)^2$, before it then continues
with a more drastic fall-off towards the JLab measurement at
$Q^2=0.9~(GeV/c)^2$. A dipole fall off of $\alpha_E$ that has been
suggested by the Bates~\cite{gp6} and the JLab~\cite{gp5,jlabgp}
data, but not supported by the MAMI~\cite{gp3,gp4} ones, is
consistent only with the lower end of the $\alpha_E$ experimental
uncertainty of our results at $0.20~(GeV/c)^2$. Another MAMI
experiment~\cite{vcsq2} has performed measurements of the electric
GP in the same momentum transfer region and its upcoming results are
expected to shed more light into the $Q^2$ dependence of $\alpha_E$.
The $\alpha_E$ polarizability has been studied within a variety of
theoretical frameworks, as exhibited in Fig.~\ref{fig:results}, and
the new measurement offers new input, and constraints, to these
calculations. The results are in agreement with the
BChPT~\cite{bchpt} and the HBChPT~\cite{gp9} calculations, while the
predictions of the NRQCM~\cite{gp15}, Effective Lagrangian
Model~\cite{gpkorchin}, and the Linear Sigma Model~\cite{gp16} tend
to underestimate the magnitude of $\alpha_E$, as also suggested by
the rest of the world data. It has to be pointed out that all
theoretical calculations predict a monotonic fall off of $\alpha_E$
and none of them is able to account for a non trivial $Q^2$
dependence that the world data suggest.

An extraction of the $CMR$ has been performed for the first time
through the measurement of the $p(e,e'p)\gamma$ reaction. The new
result is presented in Fig.~\ref{fig:results} (right panel), along
with the results from the pion channel measurements
\cite{spaplb,pos01,spaprl,stave,elsner,aznauryan,
ndeltaplb,sparepja}. One can note the excellent agreement between
the photon and the pion channel results at $Q^2=0.20~(GeV/c)^2$. The
results are compared to the phenomenological analysis of MAID
\cite{kama,mai00}, DMT \cite{dmt}, SAID \cite{said} and the
theoretical predictions of Sato-Lee \cite{sato_lee}, Capstick
\cite{capstick}, hypercentral quark model (HQM) \cite{hqm}, the
Lattice-QCD calculation \cite{dina}, the large-Nc calculation
\cite{largenc}, the DSEM~\cite{dsem}, the linear $\sigma$-model
(LSM)~\cite{lsm}, the chiral effective field theory of
Pascalutsa-Vanderhaegen (PV)~\cite{pasc} and the Gail-Hemmert (GH)
\cite{hemmert}. The results support the dominant role of the mesonic
degrees of freedom at the large distance scale and the conclusion
that approximately half of the magnitude of the Coulomb quadrupole
amplitude is attributed to the mesonic cloud at low~$Q^2$, as also
suggested by the pion channel measurements~\cite{spaplb}. The unique
aspect of this measurement is that it is the first measurement
performed through a different reaction channel and utilizes a
completely different extraction framework. This is an important step
considering that the word data for the resonant quadrupole
amplitudes are typically accompanied by a model uncertainty that is
associated with the treatment of the numerous physical background
amplitudes and is not sufficiently constrained through the
experimental measurements of the pion channel. A theoretical model
is typically utilized for the treatment of these amplitudes thus
introducing a model uncertainty to the world data results. A future
re-analysis of the same data, utilizing a new theoretical model,
could thus potentially increase the significance of these model
uncertainties. In this work, we explored the same quantity through a
different reaction mechanism and within a completely different
theoretical framework, thus providing a cross-check of the results
for the $CMR$ extracted via pion electroproduction as well as an
independent verification of the world data model uncertainties

We would like to thank the MAMI accelerator staff for their
outstanding support. This work is supported by the US Department of
Energy award DE-SC0016577, by the Federal State of
Rhineland-Palatinate, and by the German Research Foundation with the
Collaborative Research Center 1044.


\section*{References}

\bibliographystyle{elsarticle-num}

\begin{thebibliography}{99}




\bibitem{reviewpolar} F. Hagelstein, R. Miskimen and V. Pascalutsa,
Prog. Part. Nucl. Phys. 88 (2016) 29-97

\bibitem{Guichon:1998xv} P.~A.~M.~Guichon and M.~Vanderhaeghen,  Prog.\ Part.\ Nucl.\ Phys.\  {\bf 41}, 125 (1998).

\bibitem{Gorchtein:2009qq}  M.~Gorchtein, C.~Lorce, B.~Pasquini and M.~Vanderhaeghen,  Phys.\ Rev.\ Lett.\  {\bf 104}, 112001 (2010).

\bibitem{gp2} P.A.M. Guichon, G.Q. Liu and A.W. Thomas, Nucl. Phys. A 591 (1995) 606.

\bibitem{gpred1} D. Drechsel, G. Knochlein, A.Y. Korchin, A. Metz, S. Scherer, Phys.
Rev. C 57, 941 (1998)

\bibitem{gpred2} D. Drechsel, G. Knochlein, A.Y. Korchin, A. Metz, S. Scherer, Phys.
Rev. C 58, 1751 (1998)

\bibitem{gp1} V. Olmos de Leon, et al., Eur. Phys. J. A10 (2001) 207

\bibitem{rcs} M. Schumacher, Prog. Part. Nucl. Phys. 55, 567 (2005)

\bibitem{vcspaper} N.~Sparveris {\it et al.}, {\it Phys. Rev.} {\bf C78}, 018201
(2008).

\bibitem{soh} A.M. Bernstein and C.N.~Papanicolas, AIP Conf. Proc. {\bf 904}, 1 (2007).

\bibitem{gp12} B. Pasquini, M. Gorchtein, D. Drechsel, A. Metz, M. Vanderhaeghen, Eur. Phys. J. {\bf A 11}, 185-208 (2001).

\bibitem{gp13} D. Drechsel, B. Pasquini, M. Vanderhaeghen,  Phys. Rept. 378, 99-205 (2003).

\bibitem{Pasquini:2018wbl} B.~Pasquini and M.~Vanderhaeghen, Ann.\ Rev.\ Nucl.\ Part.\ Sci.\  {\bf 68}, 75 (2018)

\bibitem{gp3} J. Roche, et al., Phys. Rev. Lett. 85 (2000) 708-711.

\bibitem{gp4} P. Janssens, et al., Eur. Phys. J. A37 (2008) 1-8

\bibitem{helprog} H. Fonvieille, Prog. Part. Nucl. Phys. 55 (2005) 198-214

\bibitem{mamisetup} K.I. Blomqvist et al., Nucl. Instrum. Methods A
403, 263 (1998).

\bibitem{blombergthesis} A. Blomberg, Ph.D. thesis, Temple University, 2016.

\bibitem{mamiproposal} N.~Sparveris {\it et al.}, {\it Study of the
nucleon structure by Virtual Compton Scattering measurements at the
$\Delta$ resonance}, MAMI A1 Proposal.

\bibitem{radiative} M. Vanderhaeghen, et al., Phys. Rev. C62 (2000) 025501

\bibitem{jlabgp} H. Fonvieille, et al., Phys. Rev. C86 (2012) 015210

\bibitem{spaplb} N. F. Sparveris {\it et al.}, {\it Phys. Lett.} {\bf B651}, 102 (2007).

\bibitem{gp6} P. Bourgeois, et al., Phys. Rev. Lett. 97 (2006) 212001

\bibitem{gp5} G. Laveissiere, et al., Phys. Rev. Lett. 93 (2004) 122001

\bibitem{vcsq2} H.~Merkel et al., MAMI-A1 proposal A1-1/09~(2009),\\
URL:
http://wwwa1.kph.uni-mainz.de/A1/publications/proposals/MAMI-A1-1-09.pdf

\bibitem{bchpt} V.~Lensky, V.~Pascalutsa, M.~Vanderhaeghen, Eur. Phys. J. {\bf C 77},(2017) no.2, 119.

\bibitem{gp9} T. R. Hemmert, B. R. Holstein, G. Knochlein, D. Drechsel, Phys. Rev. D62 (2000) 014013.

\bibitem{gp15} B. Pasquini, S. Scherer, D. Drechsel,  Phys. Rev. C63 (2001) 025205.

\bibitem{gpkorchin} A. Korchin and O. Scholten, Phys. Rev. C58 (1998)
1098

\bibitem{gp16} A. Metz, D. Drechsel,  Z. Phys. A356 (1996) 351-357.


\bibitem{pos01} T. Pospischil {\it et al.}, Phys. Rev. Lett. {\bf 86}, 2959 (2001).

\bibitem{spaprl} N.F.~Sparveris {\it et al.}, Phys. Rev. Lett. {\bf
94}, 022003 (2005).

\bibitem{stave} S.~Stave {\it et al.}, Eur. Phys. J. {\bf A 30}, 471 (2006).

\bibitem{elsner} D.~Elsner {\it et al.}, Eur. Phys. J. {\bf A 27} 91-97 (2006).

\bibitem{aznauryan} I. G. Aznauryan {\it et al.}, {\it Phys. Rev.} {\bf C80}, 055203 (2009)

\bibitem{ndeltaplb} A. Blomberg {\it et al.}, {\it Phys. Lett.} {\bf B760}, 267 (2016).

\bibitem{sparepja} N.~Sparveris {\it et al.}, Eur. Phys. J. {\bf A 49}, 136 (2013).

\bibitem{kama} S.S. Kamalov {\it et al.}, Phys. Lett. {\bf B 522}, 27 (2001).

\bibitem{mai00} D.~Drechsel {\it et al.}, Nucl. Phys. {\bf A 645}, 145 (1999).

\bibitem{dmt} S.S. Kamalov and S.N. Yang, Phys. Rev. Lett. {\bf 83}, 4494
(1999)

\bibitem{said} R.A.~Arndt, {\it et al.} Phys. Rev. {\bf C66}, 055213 (2002); nucl-th/0301068 and http://gwdac.phys.gwu.edu

\bibitem{sato_lee} T.~Sato and T.-S.H.~Lee, Phys. Rev. {\bf C63}, 055201 (2001).

\bibitem{capstick} S. Capstick and G. Karl, {\it Phys. Rev.}  {\bf D41}, 2767 (1990).

\bibitem{hqm} M. De~Sanctis {\it et al.}, Nucl. Phys. {\bf A 755}, 294 (2005).

\bibitem{dina} C.~Alexandrou {\it et al.}, Phys. Rev. Lett. {\bf 94}, 021601

\bibitem{largenc} V.~Pascalutsa and M.~Vanderhaeghen, Phys. Rev. D 76, 111501
(2007).

\bibitem{dsem} Jorge Segovia, Ian C. Cloët, Craig D. Roberts and Sebastian M.
Schmidt, Few Body Syst. 55 (2014) pp. 1185-1222

\bibitem{lsm} M. Fiolhais, B. Golli, S. Sirca, {\it Phys. Lett.} {\bf B373}, 229 (1996).

\bibitem{pasc} V.~Pascalutsa and M.~Vanderhaeghen, {\it Phys. Rev.} {\bf D73}, 034003 (2006).

\bibitem{hemmert} T. A.~Gail and T. R.~Hemmert, Eur. Phys. J. {\bf A 28} (1), 91-105
(2006).















\end{thebibliography}

\end{document}